\documentclass{article}
\usepackage[utf8]{inputenc}
\usepackage[english]{babel}

\usepackage[letterpaper,top=2cm,bottom=2cm,left=3cm,right=3cm,marginparwidth=1.75cm]{geometry}
\usepackage{caption}
\usepackage{subcaption}

\usepackage{amsmath}
\usepackage{graphicx}
\usepackage[colorlinks=true, allcolors=blue]{hyperref}

\title{A subjective study of the perceptual acceptability of audio-video desynchronization in sports videos}
\author{Joshua Peter Ebenezer}

\date{}

\begin{document}

\maketitle
\begin{abstract}

This paper presents the results of a study conducted on the perceptual acceptability of audio-video desynchronization for sports videos. The study was conducted with 45 videos generated by applying 8 audio-video offsets on 5 source contents. 20 subjects participated in the study. The results show that humans are more sensitive to audio-video offset errors for speech stimuli, and the complex events that occur in sports broadcasts have higher thresholds of acceptability. This suggests the tuning of audio-video synchronization requirements in broadcasting to the content of the broadcast.
\end{abstract}

\section{Introduction}

Audio and Video of livestreamed content arrives to consumers via different transmission pathways which may be out-of-sync. Perceptible differences in the timing of the audio and video streams can cause annoyance to viewers. Prior research in the area has primarily focused on delays in human speech, i.e., when the video and audio of a human speaker do not align temporally. The International Telcommunications Union (ITU) published Rec. ITU-R BT.1359-1 addressing this issue and releasing experiments and data to establish the thresholds of perceptible A/V desynchronization for clips of newsreaders\cite{bt1359} and recommended detectability thresholds of +45 ms to -125 ms of audio leading video. However, these recommendations were made on the basis of desynchronized clips of a newsreader, and not other types of broadcasting content such as sports. 

The sensitivity of human perception to A/V desynchronization of speech according to the ITU 1359 study is shown in Fig.1. The visual cues of lip movement are easy to match with speech and hence lip-sync error detection has very low thresholds of +45 ms to -125 ms. However, there are a number of other types of content, such as sports, for which there is no data on how easy it is for humans to perceive and how tolerable it is for them. 
\begin{figure}
    \centering
    \includegraphics[width=0.8\textwidth]{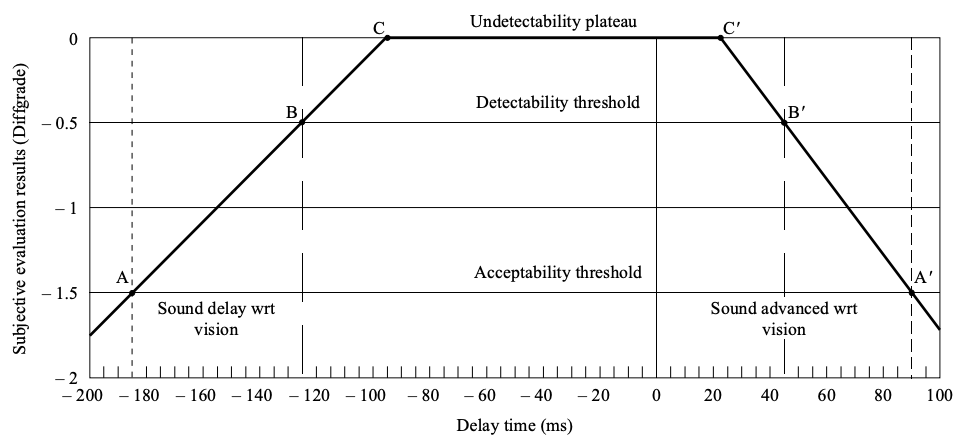}
    \caption{Sensitivity vs A/V delay for speech. Figure taken from ITU-R BT.1359-1.}
    \label{fig:my_label}
\end{figure}

\section{Prior work}

Steinmetz~\cite{steinmetz} conducted a study on perceptual thresholds for lip-sync error detection and found that offsets of $\pm 80$ ms were acceptable to observers. Vatakis and Spence~\cite{music} conducted a study on music videos that had an A/V offset and found that the thresholds for guitar and piano music were higher than for speech. Conrey and Pisoni~\cite{conrey} conducted a survey of 9 A/V synchronization studies and noted that only two of the 9 studies had non-speech stimuli, namely Lewkowicz~\cite{lewkowicz} and Dixon~\cite{dixon}, who conducted experiments with a hammer hitting a nail and a bouncing green disk, respectively. Dixon and Spitz observed that it was easier to detect synhronization errors for the the stimuli of the hammer hitting the nail than for speech.  Conrey and Pisoni also conducted their own experiment with a red disk paired with a 2000 Hz tone and found that offsets of +153 ms to -247 ms were perceivable for this artificial stimulus and that these thresholds were very similar to speech thresholds. Generally speaking, these studies show a wide range of thresholds for A/V desynchronization detection, ranging from detection thresholds of 45 ms to 250 ms.
\par 
It is is unclear what neural mechanisms underlie the influence of the synchronization of audio and video on on perception. Light travels faster than sound and hence video signals reach the brain faster than audio signals. However, there are certain calibrating mechanisms in the brain that synchronize our joint perception of sound and vision. Bushara et al~\cite{bushara} used positron emission tomography to demonstrate that the insula mediates audio-video signals at an early stage of cortical processing, and could play a role in this synchronization. Lennert et al.~\cite{lennert}  Stevenson et al.~\cite{stevenson} showed that  there are is a synchrony-defined subregion of multisensory superior temporal cortex (mSTC) that responded only when auditory and visual stimuli were synchronous, whereas a bimodal subregion of mSTC showed increasing activation with increasing levels of A/V synchronization error. Lennert et al.~\cite{lennert} used magnetoencephalography to show that the brain performs active temporal recalibration of audio and video. They proposed that visual signals are delayed in the brain to account for the slower speeds of audio signals, and that the relative delay between the video and audio signals is a dynamic process that depends on each person's recent exposure to stimuli.
\par 
To the best of the author's knowledge, there is no prior work on the perception of A/V desynchronization for sports videos. Noise from the audience, overlaid commentary, and the nature of the sport are all confounding factors that prevent conclusions drawn from simplistic and synthetic non-speech stimuli being applied to sports videos. When a sports event is livestreamed, errors can be introduced due to delays between the video camera and the microphone at the stadium, or due to denoising and quality control processing that have different latencies for video and audio. Such delays cannot be corrected immediately due to the low latency requirements of livestreaming. Sports are watched by millions of customers across the globe and such delays negatively affect their quality of experience, and therefore there is a need to study and understand the perception of audio-video desynchronization for sports. 
\section{Dataset}

Five source videos were chosen from YouTube: 1 tennis video, 1 cricket video, 1 soccer video, 1 American football video, and 1 human speech video. Each source video was used to generate 8 videos with varying A/V delays, for a total of 45 videos. The offsets were chosen differently for each of the videos due to the differing difficulties of identifying audio-visual cues from them. The ffmpeg software library was used to generate the offsets from longer videos, after which the videos were cut with ffmpeg at specified locations and for the specified durations. The offsets and durations of each video are listed in Table~\ref{tab:offsets}.

\begin{table}[]
    \centering
    \begin{tabular}{|c|c|c|}
     \hline
        \textsc{Video content} & \textsc{Duration (s)} & \textsc{A/V Offsets (s)}  \\
        \hline
        Soccer & 15 & [-1.5,-0.8,-0.4,-0.2,0,0.2,0.4,0.8,1.5] \\
        \hline 
        Cricket & 10 &  [-1,-0.8,-0.4,-0.2,0,0.2,0.4,0.8,1] \\
        \hline 
        Football & 15 & [-1.5,-0.8,-0.6,-0.2,0,0.2,0.6,0.8,1.5] \\
        \hline 
        Tennis & 10 & [-1.5,-1,-0.6,-0.2,0,0.2,0.6,1,1.5]\\
        \hline 
        Newsreader & 10  & [-0.4,-0.2,-0.1,-0.05,0,0.05,0.1,0.2,0.4]\\
        \hline 
    \end{tabular}
    \caption{Offsets and durations of each video}
    \label{tab:offsets}
\end{table}

\begin{figure}
     \centering
     \begin{subfigure}[b]{0.3\textwidth}
         \centering
         \includegraphics[width=\textwidth]{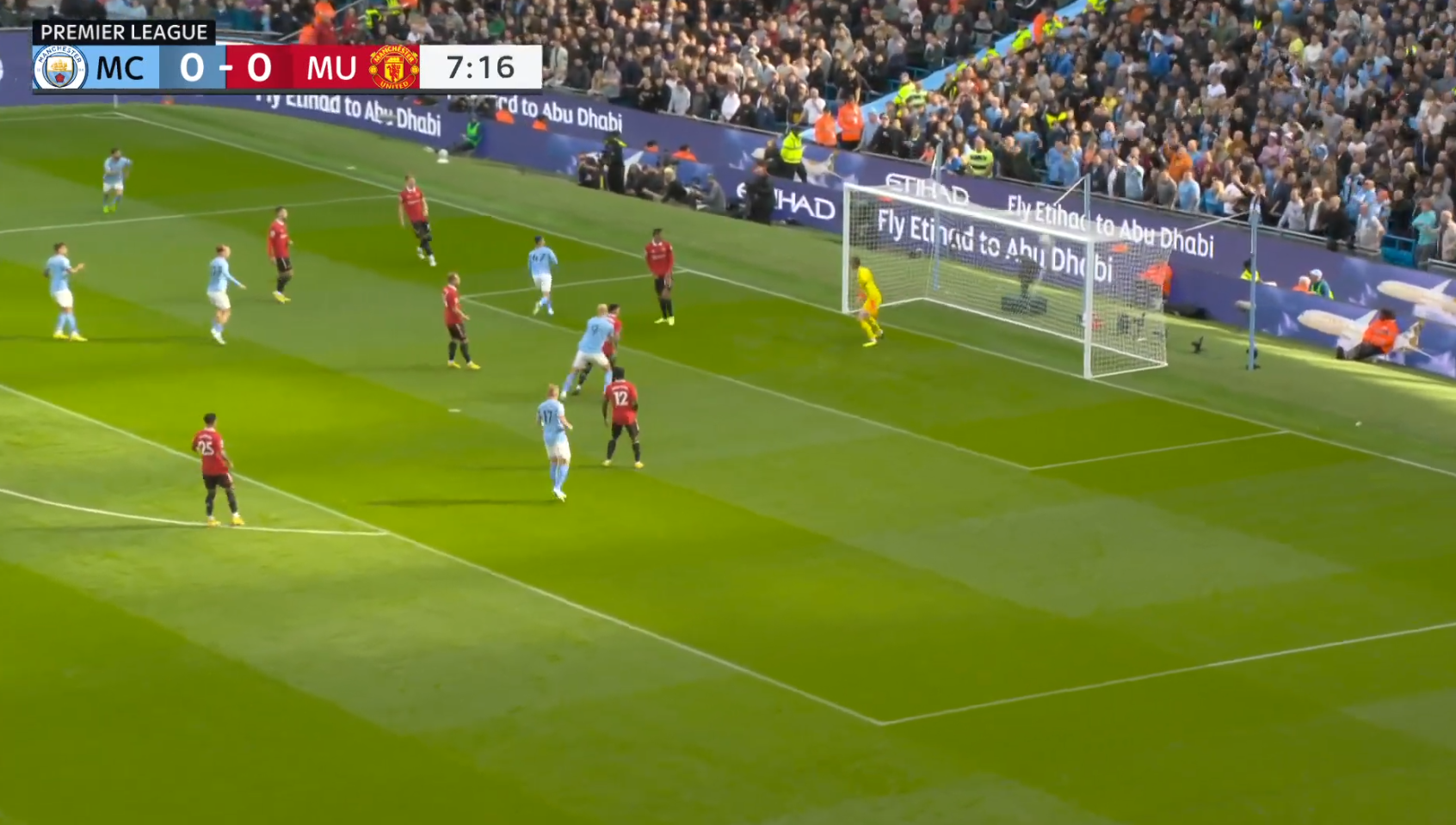}
         \caption{Soccer}
         \label{fig:soccer}
     \end{subfigure}
     \hfill
     \begin{subfigure}[b]{0.3\textwidth}
         \centering
         \includegraphics[width=\textwidth]{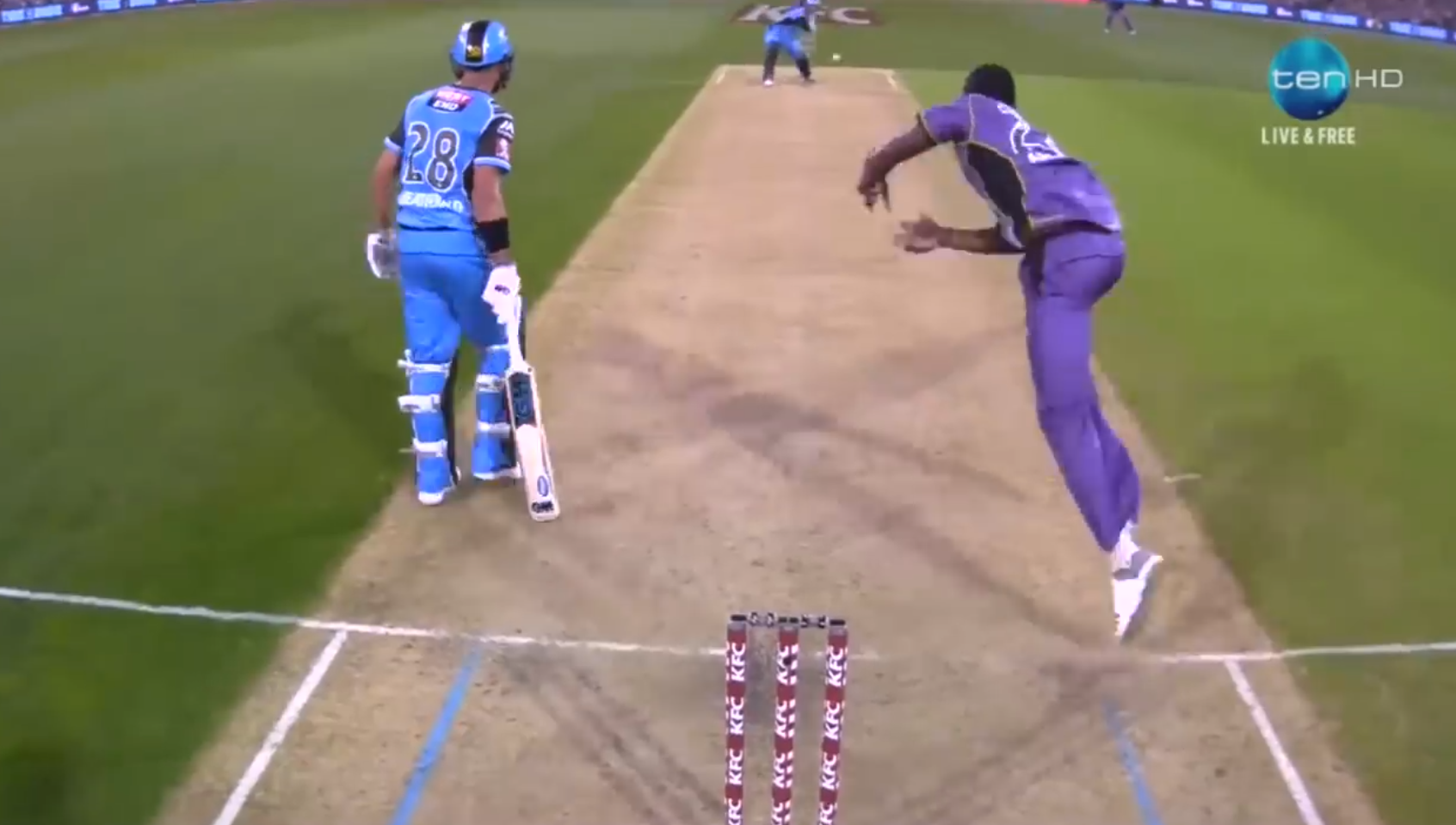}
         \caption{Cricket}
         \label{fig:cricketx}
     \end{subfigure}
     \hfill
     \begin{subfigure}[b]{0.3\textwidth}
         \centering
         \includegraphics[width=\textwidth]{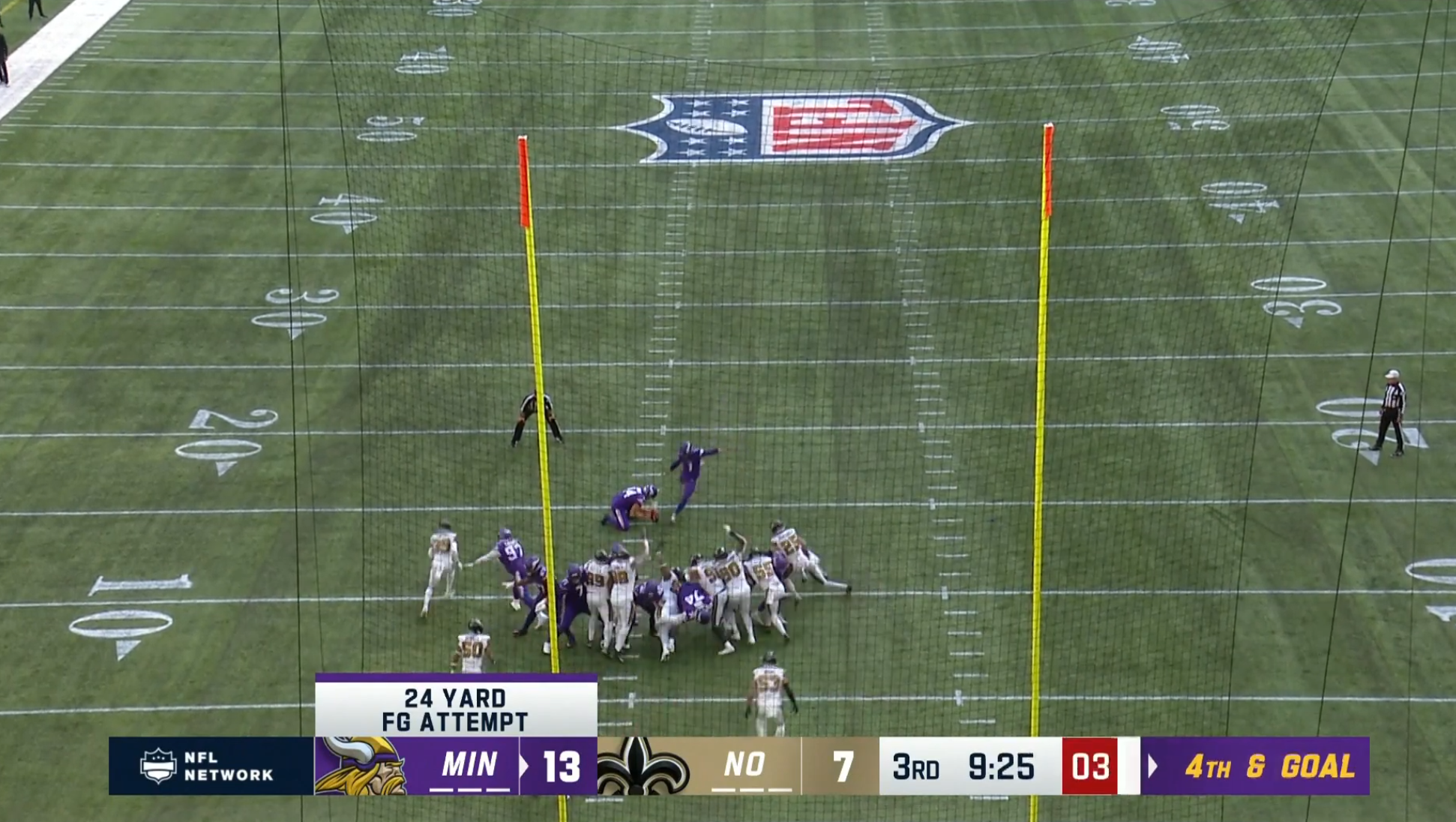}
         \caption{Football}
         \label{fig:nfl}
     \end{subfigure} \\
          \begin{subfigure}[b]{0.3\textwidth}
         \centering
         \includegraphics[width=\textwidth]{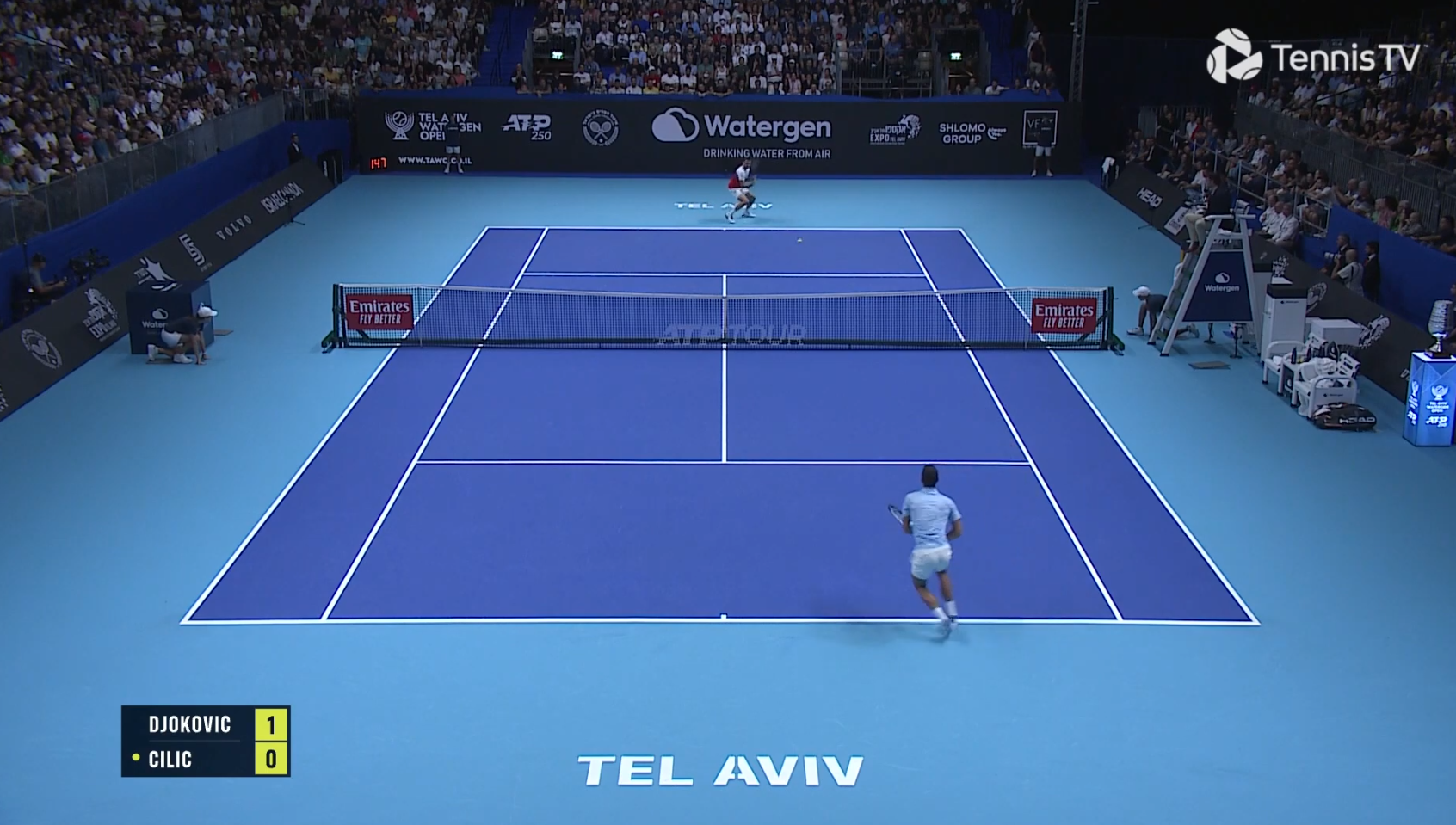}
         \caption{Tennis}
         \label{fig:tennis}
     \end{subfigure}
     \quad
     \begin{subfigure}[b]{0.3\textwidth}
         \centering
         \includegraphics[width=\textwidth]{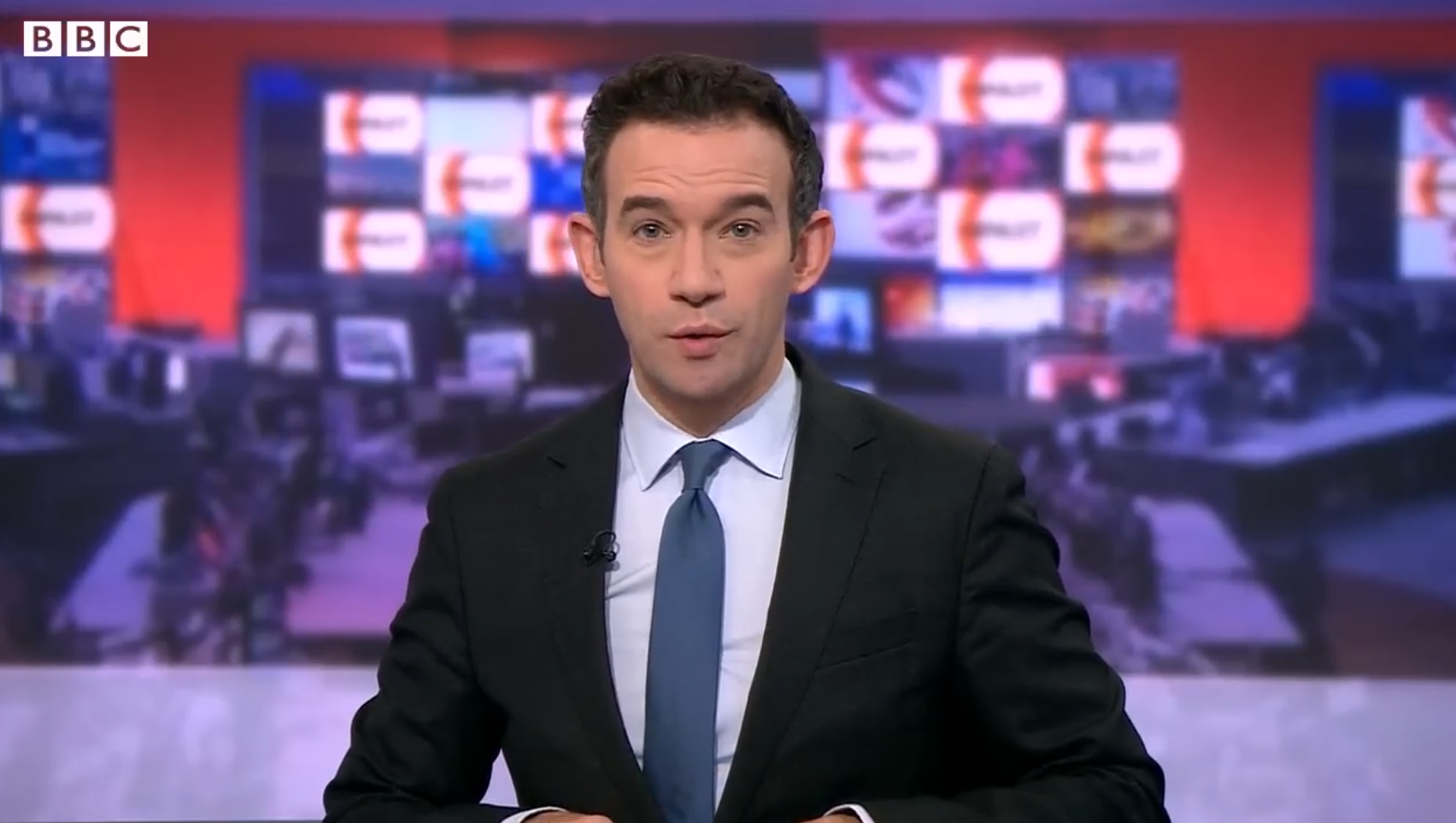}
         \caption{Newsreader}
         \label{fig:newsreader}
     \end{subfigure}
        \caption{Screenshots from the videos}
        \label{fig:screenshots}
\end{figure}

Representative screenshots of each of the stimuli are shown in Fig.~\ref{fig:screenshots}. In the soccer video, players are seen passing the ball around between themselves in the beginning of the video. The commentator names each person who receives the ball as they receive it with a slight delay that is typical of English commentating. The pace of the passing increases as the ball moves closer to the goal until one of the players receives a pass and immediately nets the ball. The crowd erupts in celebration and the commentator yells out the name of the player who scores the goal the moment the ball hits the back of the net. The atmosphere is so boisterous that one cannot hear the ball being hit. The only audio cues are therefore the commentary and the reaction of the crowd, and these have to be matched with the visual cues of the ball being passed as well as the goal being scored.  \par 
In cricket, microphones are placed on the wickets so that when the wickets are hit a loud sound is heard. The microphone also picks up the noise of the bat hitting the ball.  The cricket video begins with the bowler taking the bails of the wickets. This is followed by a replay of the incident, starting from when the bowler throws the ball to the batsman. The batsman hits the ball, which produces a loud sound, and the ball falls near the bowler. The bowler immediately runs to it, turns around, and throws the ball at the wicket, hitting it and producing another loud sound. Hence, the strongest audio cues in this video are the sounds of the bat hitting the ball and the ball hitting the wicket. The commentators talk about the incident and their audio overlays the on-field audio but is not louder than it.
\par 
At the start of the football video, the quarterback passes the ball to a receiver, who is immediately tackled to the ground by a player from the opposing team. The commentary follows the actions. The scene then cuts to a kick from the 15 yard line. The sound of the kick is audible although the commentary and the noise of the crowd mask it to some extent.
\par 
The tennis video is of a rally between Djokovic and Cilic. The play is smooth, unbroken and periodic. The ball is exchanged consistently without any stoppage. The audience and commentator are silent, as is customary of tennis matches. The only auditory and visual cues are the racquets hitting the ball.
\par 
The newsreader video is of a BBC anchor introducing a story about dog intelligence. The setting is in the studio and the lips of anchor can be visually matched to the audio of their speech. 

\section{Study}

\begin{figure}
    \centering
    \includegraphics[width=0.4\textwidth]{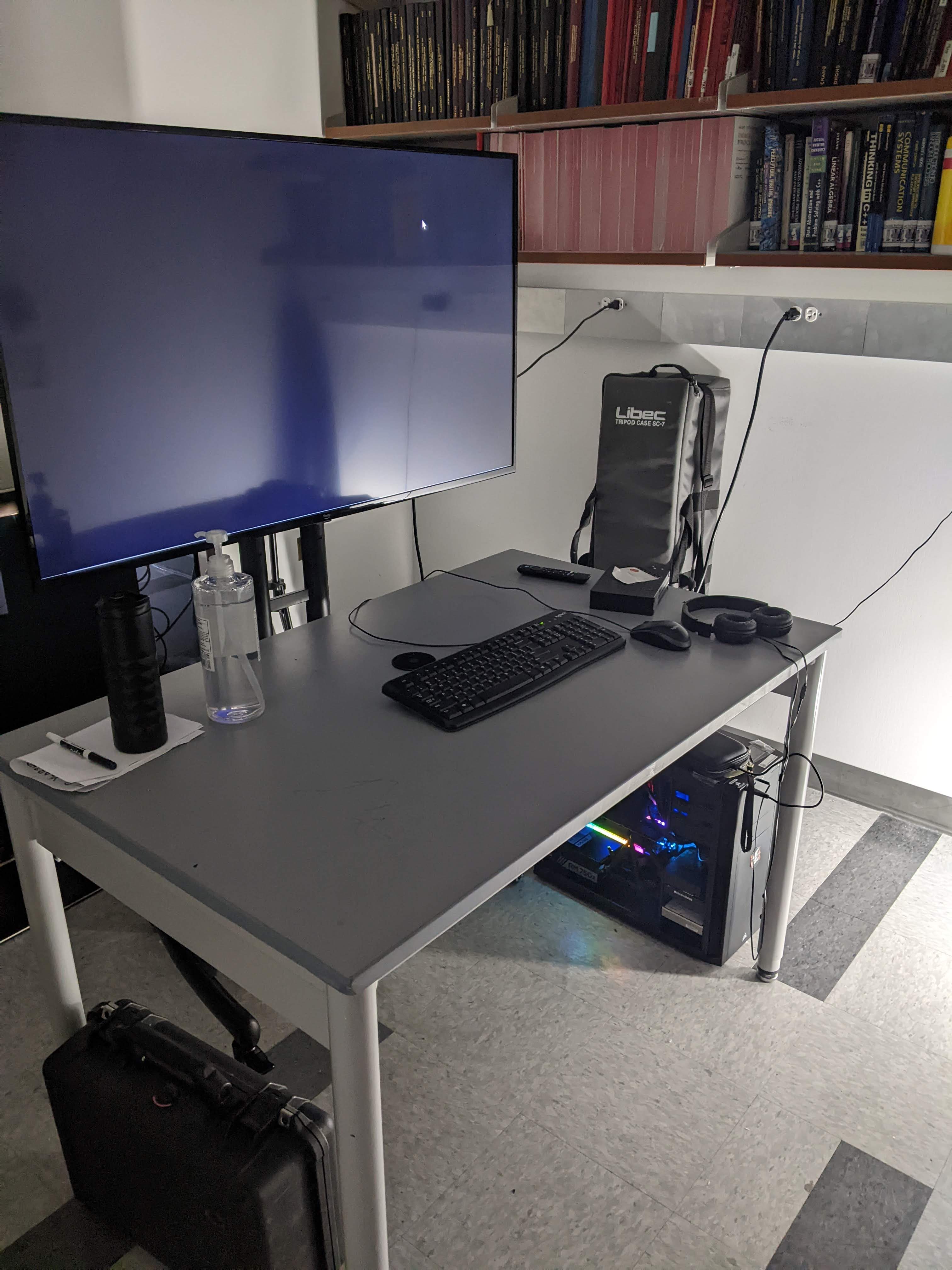}
    \caption{Photo of the study setup}
    \label{fig:setup}
\end{figure}

The study was conducted at the Department of Electrical and Computer Engineering at UT Austin. 20 students between the ages of 21 and 26 were recruited to participate in the study. An Amazon Fire TV was used as the display device. A computer with a NVIDIA 3090 Graphics card and an Intel i7 CPU was connected to the TV with a HDMI 2.1 cable and was used to power the display rendering. A JBL wired headset was used as the audio device. The videos were played on VLC media player on Windows 10. Fig.~\ref{fig:setup} shows a photo of the setup. 

\begin{figure}
     \centering
     \begin{subfigure}[b]{0.48\textwidth}
         \centering
         \includegraphics[width=\textwidth]{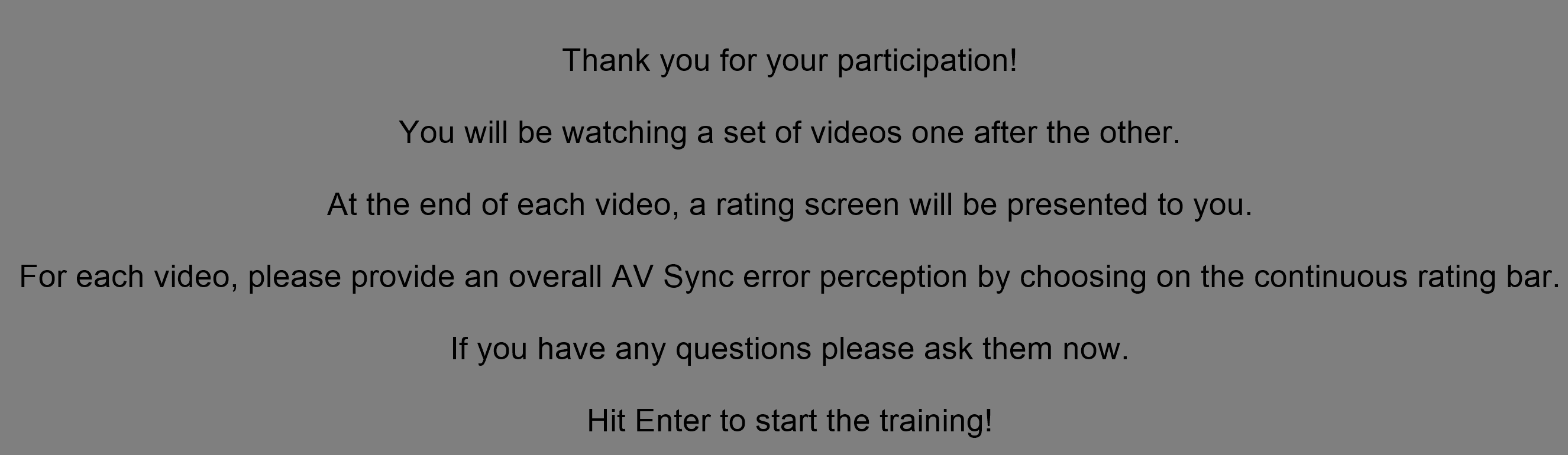}
         \caption{Instructions}
         \label{fig:instructions}
     \end{subfigure}
     \hfill
     \begin{subfigure}[b]{0.48\textwidth}
         \centering
         \includegraphics[width=\textwidth]{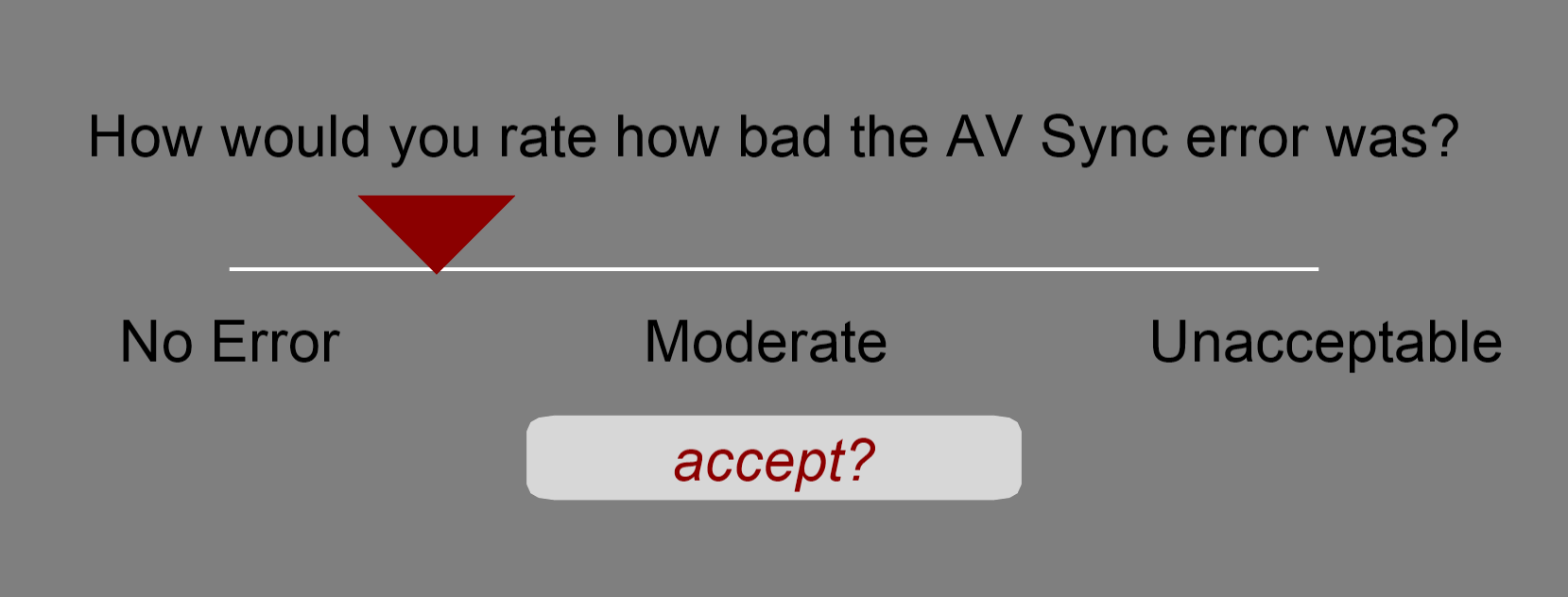}
         \caption{Scoring Screen}
         \label{fig:score}
     \end{subfigure}
             \caption{Screenshots from the scoring program}
        \label{fig:screenshots_score}
\end{figure}

The study took approximately 20 minutes to complete. Before the study began, a training session was conducted with baseball videos with differing A/V desynchronization. This was used to calibrate the subject responses, as well as to familiarize the subjects with the interface and purpose of the study. A screen with instructions was shown at the start of the training session. Instructions were also delivered verbally. Fig.~\ref{fig:instructions} shows a screenshot of the instructions and Fig.~\ref{fig:score} shows a screenshot of the scoring interface.

Following the training session, the study would begin. The videos were shown in random order, but ensuring that videos of the same content were not adjacent to each other. The order of videos was different and randomized for each subject. This removal of biases caused by the order of the videos played would counter, to some extent, the effect of the perceptual biases discovered in \cite{lennert}, as well as biases caused by fatigue. At the end of each video, a rating screen was shown with the question ``How badly would you rate the AV sync error?". Users could move a slider along a bar using their mouse. The slider had three verbal markers: ``No error", ``Moderate", and ``Unacceptable" placed at the beginning, middle, and end of the slider. The scores were recorded as numerical values from 1 to 100 with increments of 1.

\section{Score analysis}

\begin{figure}
     \centering
     \begin{subfigure}[b]{0.3\textwidth}
         \centering
         \includegraphics[width=\textwidth]{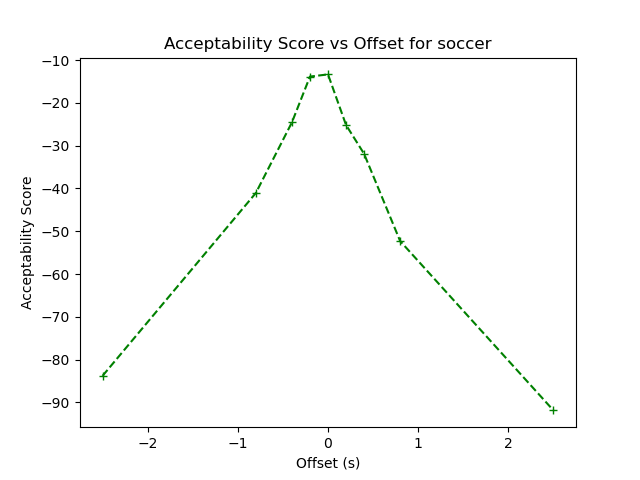}
         \caption{Soccer}
         \label{fig:sureal_soccer}
     \end{subfigure}
     \hfill
     \begin{subfigure}[b]{0.3\textwidth}
         \centering
         \includegraphics[width=\textwidth]{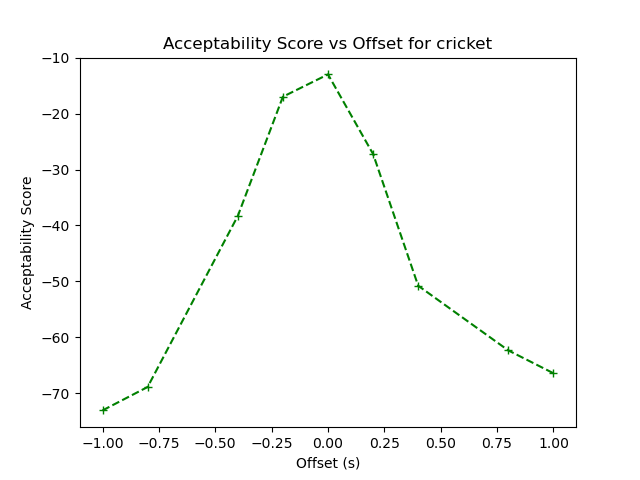}
         \caption{Cricket}
         \label{fig:sureal_cricket}
     \end{subfigure}
     \hfill
     \begin{subfigure}[b]{0.3\textwidth}
         \centering
         \includegraphics[width=\textwidth]{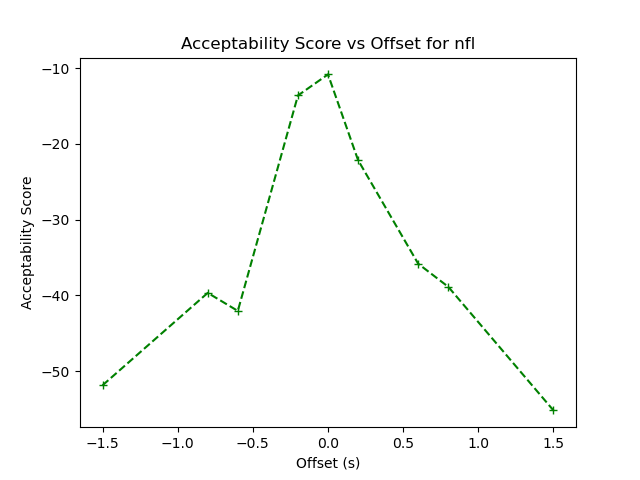}
         \caption{Football}
         \label{fig:sureal_nfl}
     \end{subfigure} \\
          \begin{subfigure}[b]{0.3\textwidth}
         \centering
         \includegraphics[width=\textwidth]{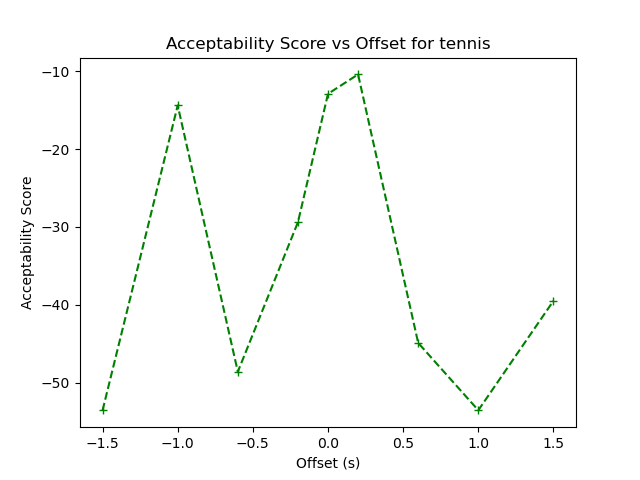}
         \caption{Tennis}
         \label{fig:sureal_tennis}
     \end{subfigure}
     \quad
     \begin{subfigure}[b]{0.3\textwidth}
         \centering
         \includegraphics[width=\textwidth]{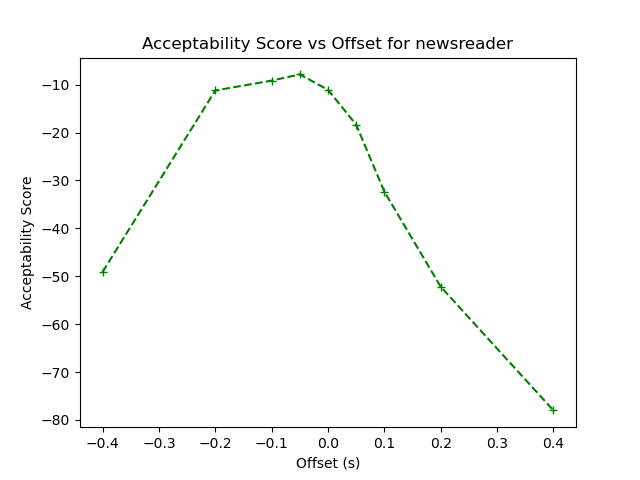}
         \caption{Newsreader}
         \label{fig:sureal_newsreader}
     \end{subfigure}
        \caption{Acceptability vs offset}
        \label{fig:sureal}
\end{figure}

The scores were considered as "acceptability" scores and aggregated using the Sureal model~\cite{sureal} of subjective data. Let $u_{ij}$ be the acceptability score of subject $i$ for video $j$. Each opinion score for video $u_{ij}$  is modelled as a sample from the random variable $U_{ij}$, defined as  
\begin{equation}
    U_{ijr} = \Psi_j + \Delta_i + \nu_i X_i 
\end{equation}

where $\Psi_j$ is the true A/V acceptability of stimulus $j$, $\Delta_i$ represents
the bias of subject $i$, the non-negative term $\nu_i$ represents the
inconsistency of subject $i$, and $X \sim N(0, 1)$ are i.i.d. Gaussian random variables. Given a collection of acceptability scores $u_{ij}$ from this experiment,  the true scores, the subject bias, and the subject inconsistency can be treated as parameters that maximize the log-likelihood of the observed scores.  This problem can be numerically solved by the Newton-Raphson method. The scores are plotted in Fig.~\ref{fig:sureal}. Here the offset is that of audio leading the video. A number of interesting observations can be made from the scores. 
\par 
Firstly, subjects are more sensitive to errors in the newsreader video than for any other clip. The acceptability score of 50 is reached for the newsreader video when the audio leads the video by 200 ms, whereas for Soccer it is reached at 1 s, for Cricket at 500 ms, for Football at 1.3 s, and for tennis at 1 s. This means that perceptual thresholds for non-speech stimuli are generally greater than for speech stimuli. Dixon and Spitz~\cite{dixon} conducted a study with a hammer and nail and found that perceptual thresholds were shorter for this stimuli than for speech. However, a hammer and nail in an isolated environment is easier to analyze than, say, a bat hitting a ball in cricket in the presence of a large and noisy crowd and overlaid commentary. The accompanying visual and auditory cues that occur in a real broadcast may contribute to a masking effect. This has a number of implications for television sports broadcasting. The allowable errors in live broadcasting need to be tuned according to studies conducted on sports videos. EBU recommendation R37~\cite{ebur37} states that the error for audio lagging video in a broadcast must be within +40 ms and -60 ms, and ITU recommends thresholds of +45 ms and -125 ms, but both of these recommendations need to be adapted to the content that is being played.
\par 
Secondly, audio leading the video is perceived as more annoying than video leading the audio. This is seen by the asymmetry of the curves. Acceptability scores are lower for the same offset in the positive direction (audio leading video) than for the negative direction (video leading audio). This is due to the fact that the brain clearly knows light to be faster than sound and, as discussed in \cite{lennert}, has certain mechanisms to correct for that. Therefore, when the sound is heard before the video, the brain has to make unexpected and unusual corrections to the signals received to synchronize them and this increases the chances of the error being detected. In addition to this, in sports videos where there is commentary, the commentary generally follows the actions and does not happen simultaneously with the action. Different commentators have different characteristic delays between the play and what they say, leading viewers to allow for greater thresholds of error for video leading audio. On the other hand, when a commentator announces something before it has happened, it is a clearly unrealistic situation that is easily detected.
\par 
The tennis scores exhibit a unique phenomenon. Due to the periodicity of the racquet hitting the ball, at certain offsets the audio of player 1 hitting the ball matches almost perfectly with the visual cue of player 2 hitting the ball, and vice versa. However, for other offsets the audio of the ball being hit occurs when the ball is in the middle of the court. The offsets for which this occurs depend on the frequency of the rally. This causes the acceptability score to fluctuate periodically.

\section{Conclusion}

The study shows that thresholds for audio-video desynchronization perception and the level of annoyance that people feel to different stimuli depends highly on the content. This data should spur the modification of existing audio-video synchrony standards for broadcasting in order to adapt to the content. Knowing these ranges for the perception and acceptability of audio-video desynchronization can also lead to the design of further experiments to evaluate the thresholds for the detection of these errors. Doing so would require generating and showing audio-visual stimuli with much smaller offsets than used in this study.  
The code and the scores from this study are being made publicly available at 
\href{https://github.com/JoshuaEbenezer/avsync_study}{https://github.com/JoshuaEbenezer/avsync\_study}.

\bibliographystyle{alpha}
\bibliography{main}

\end{document}